\documentclass[preprint,12pt]{elsarticle}

\usepackage{amssymb}
\usepackage{epsfig}
\usepackage{graphicx}
\usepackage{deluxetable}
\usepackage{mydefs}

\journal{Astroparticle Physics}

\begin{document}

\begin{frontmatter}

\title{Explaining the Cosmic-ray $e^+ / (e^- + e^+)$ and $\bar{p}/p$ Ratios using a Steady-state Injection Model}

\author[kipac]{S.-H.~Lee\corref{cor1}}
\ead{lee@yukawa.kyoto-u.ac.jp}

\author[kipac]{T.~Kamae\corref{cor2}}
\ead{kamae@slac.stanford.edu}

\author[infnpisa]{L.~Baldini}
\author[bari,infnbari]{F.~Giordano}
\author[in2p3]{M.-H.~Grondin}
\author[infnpisa]{L.~Latronico}
\author[in2p3]{M.~Lemoine-Goumard}
\author[infnpisa]{C.~Sgr\`o}
\author[kipac]{T.~Tanaka}
\author[kipac]{Y.~Uchiyama}

\cortext[cor1]{Corresponding author \\Current address: Yukawa Institute for Theoretical Physics, Kyoto University, Kitashirakawa Oiwake-Cho, Sakyo-Ku, Kyoto 606-8502, Japan}
\cortext[cor2]{Principal corresponding author}

\address[kipac]{SLAC National Accelerator Laboratory and Kavli Institute for Particle Astrophysics and Cosmology, Stanford University, Stanford, CA 94305, USA}
\address[infnpisa]{Istituto Nazionale di Fisica Nucleare, Sezione di Pisa, I-56127 Pisa, Italy}
\address[bari]{Dipartimento di Fisica ``M. Merlin" dell'Universit\`a e del Politecnico di Bari, I-70126 Bari, Italy}
\address[infnbari]{Istituto Nazionale di Fisica Nucleare, Sezione di Bari, 70126 Bari, Italy}
\address[in2p3]{Universit\'e Bordeaux 1, CNRS/IN2P3, Centre d'\'Etudes Nucl\'eaires de Bordeaux Gradignan, 33175 Gradignan, France}

\begin{abstract}
We present a model of cosmic ray (CR) injection into the Galactic space based on
recent $\gamma$-ray observations of supernova remnants (SNRs) and
pulsar wind nebulae (PWNe) by the \fermi\ Large Area Telescope (\fermi) 
and imaging atmospheric Cherenkov 
telescopes (IACTs). Steady-state injection of
nuclear particles and electrons ($e^-$) from the Galactic ensemble of SNRs,
and electrons and positrons ($e^+$) from the Galactic ensemble
of PWNe are assumed, with their injection spectra
inferred under guidance of $\gamma$-ray observations and recent development of evolution and emission
models. The ensembles of SNRs and PWNe are assumed to share the same spatial distributions. 
Assessment of possible secondary CR contribution from dense molecular clouds interacting
with SNRs is also given. CR propagation in the interstellar space is handled by
GALPROP. Different underlying source distribution models and Galaxy halo sizes are employed to estimate the systematic uncertainty of the model.
We show that this observation-based model reproduces 
the positron fraction $e^+ / (e^- + e^+)$ and antiproton-to-proton
ratio ($\bar{p}/p$) reported by \pamela\ and other previous missions
reasonably well, without calling for any speculative sources.
A discrepancy remains, however, between the total $e^-$ + $e^+$ spectrum measured by \fermi\ and our model  
below $\sim 20$~GeV, for which the potential causes are discussed.
Important quantities for Galactic
CRs including their energy injection, average lifetime in the Galaxy, and mean gas
density along their typical propagation path are also estimated.
\end{abstract}


\end{frontmatter}

\section{Introduction}
Recent observation of the positron fraction, $e^+ / (e^- + e^+)$, by \pamela\ \citep{Adriani09a} shows an excess (referred to as the $e^+$ excess, or the excess in the positron fraction)
relative to the prediction of a cosmic-ray (CR) propagation model \citep{Moskalenko98}
in the energy range between 10 and 100 GeV. The model referenced
in most analyses of the $e^+$ excess is GALPROP version 98a\footnote{We refer
to specific versions of GALPROP by the year of publication if not labeled in the literature.
A detailed description of different versions of GALPROP can be found at
http://GALPROP.stanford.edu. Various results obtained with GALPROP are
reviewed in \citet{Strong07}.}\label{footnoteGALPROP}
which assumes $e^+$ to be produced along the propagation path by the Galactic CR protons\footnote{Contribution of alpha particles and heavy ions to pion production is included in ``protons" by scaling the cross-section by an effective `nuclear enhancement factor' of $1.68$ \citep{Gaisser92}. We note that \citet{Mori09} derived a larger factor ($1.845$) for CR energy $> \sim 5$~GeV/nucleon and that the recent CR 
measurements report spectral hardening at energies $> \sim 200$~GeV/nucleon \citep{Yoon11, Adriani11b}. The present analysis is primarily concerned 
with CRs below $\sim 300$~GeV and statistically dominated by gamma-rays 
below $\sim 1$~GeV. The nuclear enhancement factor relevant to the 
present analysis should be obtained in the above CR and gamma-ray energy ranges 
and is probably smaller than $1.845$ (see \citep{Mori09, Honda11}). 
At present we assume it to be $1.68$.}\footnote{The known gamma-ray producing 
particle processes which do not go
through neutral pions (e.g., $\eta^0 \rightarrow \gamma \gamma$ and direct photon production processes) are also included but contribute less than 1~\% in the present energy range.}
having a power-law spectrum with a locally observed index of $2.75$. A constraint applied to these analyses is that
the spectrum predicted for $e^-$ + $e^+$ or leptons\footnote{We refer to $e^-$ and $e^+$
collectively as ``leptons" later in this paper. We note ``leptons" include nominally muons,
tau particles and neutrinos.}
agrees with that measured by \fermi\ \citep{Abdo09electron, Ackermann10electron} and 
the $e^-$ spectrum measured by \pamela\ \citep{Adriani11a}.
Various additional sources of $e^+$ have been proposed to account for the $e^+$ excess, including pulsars (PSRs) and pulsar wind nebulae (PWNe) \citep[e.g,][and references therein]{Yueksel09, Profumo08, Malyshev09, Gelfand09, Grasso09, Kawanaka10};
supernova remnants (SNRs) \citep[e.g.,][and references therein]{Blasi09a, Fujita09, Ahlers09}; propagation effects \citep[e.g.,][and references therein]{Katz10, Stawarz10}; and dark
matter (DM) annihilation or decay \citep[e.g.,][and references therein]{Boezio09,
Grasso09, Meade10}.  In some references, the $e^+$ excess is discussed together with a bump in the CR lepton spectrum claimed by \atic\ \citep{Chang08}.
This bump has not been observed
by \citet{Abdo09electron, Ackermann10electron} in the latest \fermi\ measurements.
Hence we will not consider the \atic\ bump in this paper.

Measurements of the antiproton-to-proton ratio ($\bar{p}/p$) have recently been extended to
$\sim 100$~GeV by \pamela\ \citep{Adriani09b, Adriani10}. The reference model used
in the analysis of the ratio is GALPROP98b which predicts $\bar{p}$ to be
produced\footnote{We note that anti-neutrons are predicted to be produced
about equally to or substantially more than $\bar{p}$ in the high energy $pp$ interaction
dependent upon the iso-spin nature of diquark pairs produced when the QCD color string
breaks \citep{Anticic10}. They decay to $\bar{p}$ with the lifetime of $\mrm{n}$. In GALPROP,
the $\bar{p}$ inclusive cross section by \citet{Tan82} has been doubled to include them.}
in the same inter-stellar matter (ISM) by the same Galactic CRs as
for $e^+$ \citep[e.g., ][and references
therein]{Moskalenko02, Strong04}. Some early measurements of the ratio
at lower energies at $\sim 10$~GeV gave higher values than the GALPROP98b
prediction, and possible contribution from the DM annihilation have
been discussed by \citet{Bergstrom99} and in references given in \citet{Moskalenko02}.
The new \pamela\ measurement agrees well with a recent GALPROP version labeled
as DC in \citet{Moskalenko02}.  Hence we will not consider possible contribution to
CR $\bar{p}$ from the DM annihilation in this paper.

In SNRs, $\gamma$, $e^-$, $e^+$, and $\bar{p}$ are produced in the $pp$ interactions
as secondary particles which can, in principle, account for the \pamela\ $e^+$
excess. However, the yields of $\pi^+$, $\pi^-$ and $\pi^0$ are predicted to be approximately
equal for proton kinetic energies greater than $\sim 10$~GeV, hence the spectrum
of $e^+$, a daughter of $\pi^+$ decay, is tightly constrained by the
observed gamma-ray spectrum of $\pi^0$ origin. We survey below SNR-based scenarios
proposed to account for the $e^+$ excess. In the work by \citet{Blasi09a},
$e^+$ are assumed to be re-accelerated to enhance the positron fraction. The lepton
spectrum will then become harder as energy increases until radiative cooling takes
over in the multi TeV energy band. All charged CRs including $\bar{p}$
will also show similar spectral hardening \citep{Blasi09b}.
\citet{Fujita09} have assumed that $e^+$ are
produced in dense clouds by nuclear CRs accelerated when the Local Bubble
exploded. The dense clouds existing at that time are assumed to have been destroyed by now.
In this scenario, we expect to find anisotropy in the arrival direction of
high energy charged CRs and in the pionic gamma-ray emissivity at local molecular
clouds; this will be tested in future \fermi\ observations.
\citet{Ahlers09} have calculated the $e^+$ spectrum at Earth based on
proton spectra deduced from the presumed pionic gamma-ray
spectra of TeV SNRs. These SNRs, however, do not represent the Galactic ensemble
of SNRs as seen in recent \fermi\ observations \citep{Abdo09w51c, Abdo10w44,
Abdo10ic443, Abdo10w28, Abdo10w49b}. \citet{Katz10} claims that the $e^+$ excess
comes out naturally if the radiative loss time is comparable to the propagation time
($\sim 10^7$ yr) for $e^+$ of energy $\sim 30$~GeV. The
authors have assumed a non-standard propagation yet to be confirmed.
\citet{Stawarz10} note that the rollover in the high energy lepton spectrum
\citep{Abdo09electron, Aharonian08a, Aharonian09} can be explained by
the Klein-Nishina effect in the radiative loss. They then note that the observed $e^+$
excess can be reproduced only if CRs propagate
through high density regions ($n > 80$~cm$^{-3}$) and secondary positrons
pass through regions where star light density is extremely high (energy density $> 300$~eV~cm$^{-3}$).

Another group of papers claim that one or a few PWNe within a few 100~pc of the solar system
can account for the $e^+$ excess \citep[e.g. ][and references
cited therein]{Yueksel09,Profumo08,Grasso09,Malyshev09}. The PWNe are
assumed to have accumulated high energy leptons over a few $\times 10$~kyr
and have released them in an interval shorter than the propagation time 
such that the highest-energy particles are
now reaching Earth. We note that local PWNe and SNRs have been considered
as possible sources of CR positrons prior to the \pamela\ experiments
\citep[e.g.,][and references therein]{Aharonian95,Atoyan95,Kobayashi04}.
According to the recent theoretical studies \citep[e.g., ][]{Gaensler06}, energetic
leptons can be injected impulsively into Galactic space only when some condition
is met in a short epoch in their life, and the probability of such a rare impulsive
injection taking place right now from one of very few nearby PWNe is low.
On the observational side,
\hess\ has found several PWNe whose ages exceed $\sim 30$~kyr as discussed by
\citet[]{Aharonian06b, Gallant07, Funk07} and in references given in \citet{Mattana09}.
\fermi\ has found GeV emission from Vela-X \citep{Abdo10vela}.
The leptons responsible for the GeV emission have energy of a few
$\times 10$~GeV and reside in the halo of the PWN adjacent to the region where
TeV emission has been found \citep{Abdo10vela,Aharonian06a}.
These observations in the GeV-TeV band as well as recent theoretical studies \citep{Gaensler06,Zhang08,Tanaka10,Bucciantini10} suggest CR leptons are released slowly
from PWNe allowing a larger ensemble of Galactic PWNe than those within
a few 100~pc of Earth to contribute to the CR lepton spectrum.

\citet{Grasso09, Malyshev09} considered a distribution of pulsars and PWNe
within a few kpc of the solar system to be responsible for the \pamela\ $e^+$
excess. \citet{Grasso09} have fitted the positron fraction well by assuming an $e^-$
spectrum softer than that used in GALPROPv44\_500180, the reference widely
used in this kind of analyses \citep{Strong04}. In this scenario, the $e^+$ flux
at higher energies are dominated by unidentified nearby PWNe.
Recently \citet{Delahaye10} studied possible energy ranges that
nearby PSRs, PWNe, and SNRs contributed to the CR electrons and positrons
at Earth. Their contributions are predicted to give a flat positron fraction at energies
greater than 10~GeV.

\citet{Ioka08} has associated the $e^+$ excess to an impulsive injection of $e^+$
by a presumed historic GRB that created the Local Bubble. Possible
association of the Local Bubble with a GRB has been suggested earlier
\citep[e.g., ][]{Perrot03,Lallement03}: \citet{Perrot03} have predicted the CR
$e^-$ spectrum at Earth will be be harder at higher energies ($E>100$~GeV)
in such a scenario and emissivities of gamma-rays
at local molecular clouds will show directional dependence
at the $\sim 50$~\% level. Future \fermi\ observations of gamma-ray emission
from local clouds will detect such a large anisotropy if it exists.

Numerous publications have attempted to associate the $e^+$ excess with annihilation and/or
decay of dark matter particles \citep[e.g., ][and references therein]{Boezio09, 
Grasso09, Meade10}, in which a number of interesting possibilities are proposed. In this work, however,
we will not discuss any dark matter related scenarios.

All publications surveyed above attempted to reproduce the $e^+$ excess by 
interpreting a number of known local objects as CR sources using adjust-to-fit CR injection
spectra and/or adopting non-standard CR propagation processes. Although we cannot rule out all such
possibilities, it is important to study how recent gamma-ray observations constrain
the positron fraction and the $\bar{p}/p$ ratio within the conventional framework.
Another important issue with these publications is that spectra of $\gamma$, $p$, $e^-$, 
$e^+$ and $\bar{p}$ are analyzed more-or-less independently. If these CRs come
from SNRs and PWNe, or are produced in nuclear interactions with molecular clouds 
and ISM gas, there will be strong correlations among their spectra. In particular, 
the CR $e^-$ spectrum should not be treated independently of the CR $p$, $e^+$ 
and $\bar{p}$ spectra. We put these data on the coherent platform of GALPROP 
by assuming the Galactic CRs are in their steady states. 

This work focuses on the Galactic CR ratios ($e^+$ fraction and $\bar{p}/p$ ratio) 
and use published gamma-ray observation results as constraints and/or consistency check 
for our calculations. All Galactic CRs are assumed to be injected by the ensemble of SNRs and PWNe, or produced 
through their interactions with the interstellar gas in the Galaxy, or from clouds interacting with SNRs. The CR injection processes     
are all assumed to be in steady-state. 
The adopted injection spectrum of protons from SNRs is constrained by the available observations of CR proton flux at Earth,
and is similar to those assumed in recent \fermi\ analyses of the diffuse Galactic gamma-ray emission \citep{Abdo09localh1,Abdo09noexcess,Fermithirdquad}.
The injection spectrum of $e^+$ and $e^-$ from PWNe is deduced from currently available gamma-ray observations 
of PWNe by \fermi\ and \hess , guided by recent theoretical spectral evolution models. 
The CR propagation process is calculated within the robust GALPROP framework. 
The calculated local CR fluxes are renormalized to those observed at Earth, which determines the CR injection luminosities. 
We purposely remain blind to 
the measured positron fraction as well as the measured $\bar{p}/p$ ratio until the calculated results are 
compared with the measurements.

The major assumptions made in our model are summarized below:
\begin{enumerate}
\item CRs are injected from the Galactic ensembles of SNRs (primary particles: $p$, $e^-$; secondary particles $e^-$, $e^+$,
$\bar{p}$) and PWNe (primary particles: $e^-$, $e^+$).
\item The primary $p$ and $e^-$ injected from the Galactic ensemble of SNRs 
have a common steady-state spectral shape with a fixed $e/p$ ratio, except near 
the maximum energy where the $e^-$ injection spectrum is corrected 
for radiative energy loss.
\item The primary $e^-$ and $e^+$ injected from the Galactic ensemble of PWNe have a common
steady-state spectral shape and injection luminosity.
\item The primary $p$ spectrum injected from SNRs is constrained by the observed CR proton spectrum at Earth. 
\item The secondary CR spectra injected from SNRs interacting with molecular clouds are calculated assuming the $p$ injection spectrum obtained in (4) at the interacting sites. 
We use the $pp \rightarrow e^+, e^-$ cross-sections modeled by 
\citet{Kamae05,Kamae06} and the $pp \rightarrow \bar{p}$ cross-section from \citet{Tan82,Tan83a,Tan83b}.
\item The primary CR spectrum injected from PWNe is deduced from gamma-ray 
observations by \fermi\ and ACTs, and PWN evolution models, e.g. by \citet{Tanaka10} and \citet{Zhang08}.
\item SNRs and PWNe are distributed identically and continuously in the Galaxy using 
the parametric recipe implemented in GALPROP. 
Two different sets of parameters are used to test the robustness of the calculated results. 
\item The propagation, interaction and energy loss of CRs from the sources to Earth are calculated using GALPROP 
with three different Galaxy halo heights (2, 4 and 10~kpc). 
Propagation parameters in the GALPROP input, such as the diffusion coefficient and re-acceleration strength (Alfv\'{e}n velocity),
are adjusted to reproduce the observed CR proton and lepton spectra, and the B/C ratio at Earth. 
\end{enumerate}

This paper is organized as follows: We describe the three CR source classes 
assumed here in section~\ref{simulations}.
The two spatial distribution models of SNR/PWN in the Galaxy
are described in section~\ref{snrpwndistribution}.
CR injection from SNRs and PWNe are discussed in sections~\ref{SNRinjection}
and \ref{PWNinjection}, respectively.
In section~\ref{results}, the calculated positron fraction and the $\bar{p}/p$ ratio
are compared with observations and discussion is given thereon.
We also present quantitative calculations on energetics and live-times of Galactic CRs 
in the section.
Finally, the results are concluded in section~\ref{conclusions} with some comments on future prospects.

\section{Three Classes of Cosmic-ray Sources in the Galaxy}
\label{simulations}

The model consists of three classes of sources, each responsible for 
injecting a specific set of CR species into the interstellar space. 
We use GALPROP, whenever possible, 
to calculate CR injection and propagation. 
In some special cases, however, these are calculated using a simple 
propagation program specifically developed to accommodate, for example, 
non-power-law injection spectra and 
source distribution without cylindrical symmetry with respect to the Galactic Center. This will be discussed 
in more details in the relevant sections.

The three classes of CR injection sources include:
\begin{description}
\item[`SNR-propagation':] This class includes the primary CRs ($p$ and $e^-$) injected from 
the Galactic ensemble of SNRs, the secondary CRs ($e^-$, $e^+$, $\bar{p}$)
and diffuse gamma-rays ($\pi^0$ decay, bremsstrahlung and IC)
produced along the propagation path of the primaries. 
\item[`PWN-propagation':] This includes the primary $e^-$ and $e^+$ injected from 
the Galactic ensemble of PWNe, and the diffuse gamma-rays produced along 
their propagation path via bremsstrahlung and inverse Compton scattering (IC). 
\item[`SNR-cloud interaction':] We assume that a fraction of CRs accelerated at SNRs (especially those in their middle-age) are interacting with dense local clouds 
(e.g., W51C, W44, IC~443, and W28). These interaction sites contribute 
through hadronic interactions to the injection of secondary particles, 
including $e^{+/-}$, $\bar{p}$, and gamma-rays. 
Propagation of $e^{+/-}$ and $\bar{p}$ from this source class has been calculated with a 
simple propagation code described in Appendix.
\end{description}

We use the conventional 2-dimensional propagation mode in GALPROP, for which the Galaxy boundary 
possesses cylindrical symmetry; we adopt a propagation volume bounded by
$R_\mrm{max} = 30$~kpc and $Z\mrm{_{max} = 4}$~kpc (the halo height) for our \textbf{`Default'} model. 
In addition, we also construct three alternative models to check the robustness 
of our results to the assumptions made for the source distribution and the Galaxy halo size. The first alternative model adopts a different parameter set for the spatial distribution, 
while the second and third have  
the halo height modified to a smaller $Z\mrm{_{max}} = 2$ and a larger $Z\mrm{_{max}} = 10$~kpc respectively.  
We note that our results are checked to be insensitive to the change 
of the radial boundary $R_\mrm{max}$ by $\pm 10$~kpc.
Propagation parameters such as the diffusion coefficient are adjusted for models with different halo heights
in order to reproduce the observed B/C ratio. 

We also estimate, in the context of steady-state CR injection in the Milky Way, 
the significance of the injection of CR positrons from nearby sources (within a few
hundred parsec from Earth) to the reproduction of the \pamela\ positron fraction. 
Treatment of this CR contribution by these `local' sources in the steady-state regime 
will be discussed in section~\ref{results}.

\section{Spatial Distribution of SNRs and PWNe}
\label{snrpwndistribution}

\begin{figure}
\centering
\includegraphics[width=10cm]{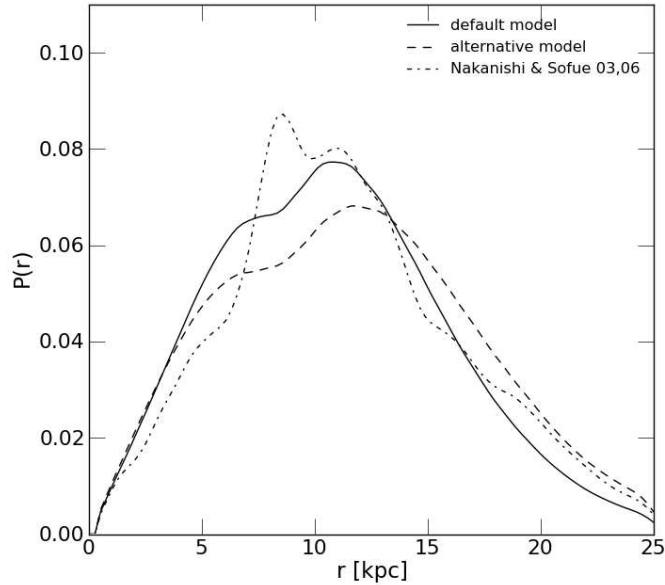}
\caption[Projected probability density for the source distribution models] 
{Probability density of source distribution plotted as a function of distance from Earth:
the parametric model in GALPROP with the default (solid) and alternative model (dash) 
parameters. Corresponding curve derived from the 3D gas distribution obtained by \citet{Nakanishi03,Nakanishi06} is
also shown as a dash-dotted line. 
}
\label{figsourcedistr}
\end{figure}

We study two spatial distribution models of CR source, parameterized in the form of: 
\begin{equation}
P(R,Z) \propto \left( \frac{R}{R_{\odot}}\right) ^{\alpha} \times \mrm{exp}\left( -\beta \left( \frac{R-R_{\odot}}{R_{\odot}} \right) \right) \times \mrm{exp}\left( -\frac{|Z|}{0.2}\right)
\end{equation}
Respectively, $R$ and $Z$ are the Galacto-centric radius and the distance from the Galactic Plane in kpc; $R_{\odot} = 8.5$~kpc. The default model adopts 
$\alpha$, $\beta = 1.25$, $3.56$ based on the pulsar distribution \citep{Strong04}, while 
the alternative model fits to a SNR distribution model described in \citet{Case98} and uses
$\alpha$, $\beta = 1.69$, $3.33$.
The two source distribution models are shown in Fig.~\ref{figsourcedistr} 
as a function of distance from Earth. Also overlaid in the figure for reference purposes 
is a corresponding curve derived from the 3D gas distribution 
obtained by \citet{Nakanishi03,Nakanishi06}. 

\section{Cosmic Ray Injection from Supernova Remnants}
\label{SNRinjection}

\subsection{Injection of Protons and Primary Electrons}
\label{SNRs}

The standard theory of particle acceleration at SNRs through diffusive shock acceleration (DSA) predicts
that the accelerated particles share a common non-thermal spectral shape when gas density is not high 
($n <$ a few cm$^{-3}$), apart from a high-energy cutoff or rollover for electrons around their maximum energy 
limited by radiative loss \citep[e.g.,][]{Blasi05,Ellison07}.
As a result, we adopt a common spectral shape for the SNR-injected protons and primary electrons, 
except near the maximum energy where radiative loss introduces a spectral cutoff for the $e^-$ spectrum. We note that little is known about the cutoff at the highest energy 
theoretically and should be readjusted when observational data become available.

\begin{figure}
\centering
\includegraphics[width=10cm]{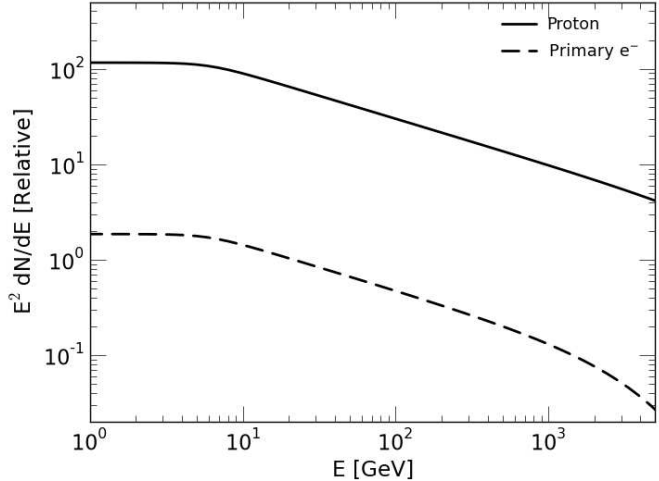}
\caption[Assumed steady-state injection spectra of protons and electrons from SNRs]
{The assumed steady-state injection spectrum of protons (solid)
and primary electrons (dashed) for the `SNR-propagation' source class. An 
exponential cut-off at the maximum energy estimated for
an age of $50$~kyr is applied to the electron spectrum shown.}
\label{figsnrprotonspectra}
\end{figure}

We use a broken power-law injection spectrum for the SNR-injected primary CRs. The spectral parameters and injection 
luminosity of protons are chosen such that the propagated spectrum fits well with the observed CR proton flux at Earth.  
The e/p ratio, K$_{ep}$, is chosen such that the calculated total $e^- + e^+$ flux, 
which are mainly contributed by the `SNR-propagation' and `PWN-propagation' classes, reproduce the 
observed CR lepton spectrum measured by \fermi\ 
\citep{Abdo09electron,Ackermann10electron} in the $50-150$~GeV energy band
(more will be discussed in section~\ref{PWNinjection}).
The adopted steady-state proton and electron injection spectrum are shown in Fig.~\ref{figsnrprotonspectra} by a thick solid line and dash-dotted line respectively,
and the relevant parameters are summarized in Table~\ref{modelparamtable1}. 

\begin{deluxetable}{lll}
\tabletypesize{\footnotesize}
\tablecolumns{3}
\tablewidth{0pt}
\tablecaption{Summary of Model Parameters}
\tablehead{\colhead{Parameter} & \colhead{Value} & \colhead{Description}}
\startdata
\sidehead{~~GALPROP Inputs}
\hline
\\
$B$ &	$8 \times \mrm{exp} \left(\frac{R_{\odot}-R}{50} \right) \times \mrm{exp} \left( -\frac{|Z|}{3} \right)\ \mu$G&	 B-field model in ISM \tna  \\
$Z_\mrm{max}$ & $2,\ 4,\ 10$~kpc & Galactic halo height \\
$D$ & $(2.9,\ 5.8,\ 10.0) \times 10^{28} \beta\ (\frac{\mathrm{R}}{4\mathrm{GV}})^{0.33}\ \mathrm{cm}^2\mathrm{s}^{-1}$ & Diffusion coefficient \tnb \\ 
$v_A$ & $30\ \mathrm{km}\ \mathrm{s}^{-1}$ & Alfv\'{e}n wave velocity in ISM \\
\\
\hline
\sidehead{~~Galactic Ensemble of Supernova Remnants \tnc}
\hline
\\
$L_{p,~>0.1\mrm{GeV}}$ & $6.42 \times 10^{40}\ \mathrm{erg}\ \mathrm{s}^{-1}$ & Total luminosity of proton \\
$L_{e,~>0.1\mrm{GeV}}$ & $1.23 \times 10^{39}\ \mathrm{erg}\ \mathrm{s}^{-1}$ & Total luminosity of primary $e^-$ \\
$K_{ep}$ & $0.02$ & The e/p ratio \\
$E_b$ & $7$~GeV & Spectral break energy \\
$\gamma_1$ & $2.0$ & Spectral index below $E_b$ \\
$\gamma_2$ & $2.5$ & Spectral index above $E_b$ \\
$n \times W_{p,~>1\mrm{GeV}}$ & $5 \times 10^{51}$ erg~cm$^{-3}$ & See footnote~$a$ in Table~\ref{snrtable} \\
\\
\hline
\sidehead{~~Galactic Ensemble of Pulsar Wind Nebulae \tnc\, \tnd}
\hline
\\
$L_{e,~>0.1\mrm{GeV}}$ & $(2.38-5.72) \times 10^{37}\ \mathrm{erg}\ \mathrm{s}^{-1}$ & Total luminosity of $e^+ + e^-$ \\
$E_b$ & $100-500$~GeV & Spectral break energy \\
$\gamma_1$ & $1.0-1.7$ & Spectral index below $E_b$ \\
$\gamma_2$ & $2.8-3.3$ & Spectral index above $E_b$ \\
\enddata
\tablenotetext{a}{$R$ and $Z$ are Galactocentric distances in kpc.}
\tablenotetext{b}{Parameter values for $Z_\mrm{max}=2,\ 4\ \mrm{and}\ 10$~kpc, respectively.}
\tablenotetext{c}{Quoted injection luminosities and $K_{ep}$ correspond to the default setup with $Z\mrm{_{max}}=4$~kpc,
$R\mrm{_{max}}=30$~kpc, and the default spatial source distribution.} 
\tablenotetext{d}{Ranges of spectral parameters given represent
the sampled possible intervals, which are assumed basing on gamma-ray observations of PWNe. 
The injection luminosity adjusts accordingly with the spectrum, with its out-coming range listed here.} 
\label{modelparamtable1}
\end{deluxetable}

According to \citet{Yamazaki06}, the maximum 
electron energy is limited to $\sim 14$~TeV for SNRs older than 1~kyr, and
$\sim 7$~TeV for those 10~kyr old; it then drops to $\sim 0.5$~TeV 
for ages of a few $\times\ 100$~kyr.
We calculate the loss-limited E$\mrm{_{max}}$ in accordance with \citet{Yamazaki06} 
(eqn. 2 and 10), assuming B$_d = 10~\mu$G, $r = 4$, $\mrm{h = E_{51} = v_{i,9} = n_0 = 1}$, 
and a hydro solution for a typical SNR lifetime of $50$~kyr within the Sedov phase, such that:

\begin{equation}
E_\mrm{max} = 13.8~\mrm{TeV} \times \left( \frac{t_\mrm{age}}{10~\mrm{kyr}} \right) ^{-\frac{3}{5}} \times \left( \frac{B_d}{10~\mu \mrm{G}} \right) ^{-\frac{1}{2}}
\label{eqn_emax}
\end{equation}

\noindent Beyond $50$~kyr, the electron injection luminosity is 
expected to be too low to influence the time-averaged injection; hence, it is neglected in this study. 
The evolution of $E_\mrm{max}$ of the continuously injected $e^-$ from the Galactic ensemble of SNRs results in a 
smooth rollover of the calculated local $e^-$ 
spectrum from the `SNR-propagation' class at around $1$~TeV. To determine the shape of the rollover, we use the simple
propagation code described in the Appendix. This rollover is then applied to the propagated primary $e^-$ spectrum first
calculated using GALPROP without the incorporation of $E_\mrm{max}$. More details of the calculation
of the rollover shape is given in the Appendix.

\subsection{Secondary Particles Contribution from SNRs}\label{Secondary}

Secondary CRs are produced when the shock-accelerated nuclei interact with the surrounding gas.  
\citet{Blasi09a} proposed that secondary positrons produced (and re-accelerated) in old SNRs may be responsible for 
the rise of the positron fraction observed by \pamela.
Meanwhile, antiprotons should also be produced and accelerated through an identical mechanism
\citep{Blasi09b}. After the first year of science operation of the \fermi\ mission, 
high-energy gamma-ray emission have been detected with high statistical significance
from 6 Galactic SNRs, and all of them have TeV counterparts observed by ground-based ACTs.
Among them, 5 are middle-aged SNRs known to be interacting with local molecular clouds.  
Their gamma-ray spectra are best explained by the decay of $\pi^0$ mesons
\citep{Abdo09w51c,Abdo10w44,Abdo10ic443,Abdo10w28,Abdo10w49b}, 
with the underlying proton spectra at the interaction sites found to follow broken power-laws. 
A summary of the derived proton spectra for these gamma-ray SNRs is given in Table~\ref{snrtable}.
The gamma-ray fluxes are found to be high, probably implying that the secondary CRs 
produced and injected by these interaction sites can be appreciable.
However, the post-break indices of the deduced broken power-law proton spectra are 
generally soft ($>2.7$), and particle acceleration in these interaction sites is probably inefficient 
beyond the break energies found around $\sim 10$~GeV (see \citet{Malkov10} 
for possible theoretical interpretation).  
We try to estimate in our model
the contribution of secondary CRs from these interaction sites to the local fluxes (the `SNR-cloud interaction' class), 
and deduce an upper-limit for the total number of these sites in our Galaxy using constraints from updated CR measurements.

\begin{deluxetable}{lccccccc}
\tablecolumns{6}
\tablewidth{0pt}
\tablecaption{SNR-cloud interaction sites observed by \fermi\ and ACTs}
\tabletypesize{\footnotesize}
\tablehead{
\colhead{SNR} & \colhead{Age} & \colhead{Distance} & \colhead{Gas dens $\times$ CR \tna}
& \multicolumn{3}{c}{PL or Broken-PL \tnb} & \colhead{Refs. \tnc} \\\\\cline{5-7}
\multicolumn{8}{c}{}\\
\colhead{Name}	& \colhead{kyr}	& \colhead{kpc} & \colhead{$n\times W_p$~[erg~cm$^{-3}$]} &
\colhead{$E_{br}$~[GeV]} & \colhead{$\alpha_{LE}$} & \colhead{$\alpha_{HE}$}
& \colhead{} \\
}
\startdata
W49B & $4.0$ & $7.5$ & $1.1 \times 10^{52}$ & $4.0$ & $2.0$ & $2.7$ & \tnd \\ 
W44 & $20$ & $3.0$ & $6 \times 10^{51}$ & $8.0$ & $1.74$ & $3.7$ & \tne  \\
W51C & $30$ & $6.0$ & $5.2 \times 10^{51}$ & $15$ & $1.5$ & $2.9$ & \tnf \\
IC~443 & $30$ & $1.5$ & $6.7 \times 10^{50}$ & $69$ & $2.09$ & $2.87$ & \tng \\
W28 (North) & $40$ & $2.0$ & $1.3 \times 10^{51}$ & $2.0$ & $1.70$ & $2.70$ & \tnh \\
\enddata
\label{snrtable}
\tablenotetext{a}{Product of the average gas density and the total kinetic 
energy of CR protons in the SNR-cloud interaction site, assuming that the observed gamma-rays luminosities 
have a pion-decay dominated origin. Errors associated 
with these estimations are likely to be within a factor less than $10$ above and below.}
\tablenotetext{b}{Spectrum of protons interacting with dense clouds at SNRs for hadronic scenarios. 
PL stands for a power-law spectrum.}
\tablenotetext{c}{References for ages and distances are found 
in the literatures cited below.}
\tablenotetext{d}{\citet{Abdo10w49b}}
\tablenotetext{e}{\citet{Abdo10w44}}
\tablenotetext{f}{\citet{Abdo09w51c,Fiasson09}}
\tablenotetext{g}{\citet{Abdo10ic443,Albert07a,Acciari09}}
\tablenotetext{h}{\citet{Abdo10w28,Aharonian08b}}
\end{deluxetable}

The injection luminosity of these secondary CRs are determined by the underlying proton spectrum and the product 
$n \times W_p$, where $n$ and $W_p$ are the average gas density and the total kinetic energy of CR protons, within each 
interaction site. With only a relatively small sample of this source class available, 
our knowledge on these quantities is very limited.
With reference to gamma-ray observations of SNRs interacting with clouds (Table~\ref{snrtable}), we adopt a typical 
$n\times W_p$ ($T_p > 1$~GeV) of $5 \times 10^{51}$~erg~cm$^{-3}$ for each site.
For this 'SNR-cloud interaction' source class, we use the same spatial distribution (Section~\ref{snrpwndistribution}), 
underlying proton spectrum, and propagation parameters as those adopted for the `SNR-propagation' source class. 
Instead of using GALPROP, we use the simple propagation code described
in the Appendix to propagate the secondary particles produced, such that the injection luminosities per interaction 
site and the non-power-law injection spectra can be accommodated.  
As has been noted before, neither re-acceleration effect nor hadronic interaction along the CR propagation path 
are considered for these secondary CRs.  

\section{Spectra of Electron and Positron Injected by PWNe}\label{PWNe}
\label{PWNinjection}

Electrons and positrons are accelerated in the magnetosphere of a pulsar, injected to
its associated PWN, and then accelerated again, presumably through diffusive shock 
acceleration across a relativistic termination shock. Electromagnetic spectra in the keV and MeV bands are mostly believed 
to originate from synchrotron radiation by
the accelerated leptons, while those in the GeV and TeV bands are dominated by 
IC scattering of the synchrotron as well as the background photon fields by the same lepton population. 
Until recently, the Crab nebula was the
only PWN detected clearly in both the GeV and TeV bands, and little was known about the
evolution of gamma-ray emission from PWNe. In the past few years, however, \hess\ has 
detected several PWNe \citep{Aharonian06b, Gallant07, Funk07}, shedding light to
the evolution of their broadband emission \citep{deJager08,Mattana09}.   
\citet{Mattana09} list 14 TeV PWNe of which \fermi\ has firmly detected GeV emission 
from Crab, Vela~X and MSH~15-52 \citep{Abdo10crab,Abdo10vela,Abdo10msh}.
Of particular importance is that \hess\ has detected TeV emission from 3 PWNe whose 
ages are estimated to be greater than $50$~kyr. This suggests
that PWNe retain TeV leptons up to or longer than $\sim 10^5$~yr.

A number of recent studies have attempted quite successfully to
construct dynamical evolution models of non-thermal emission of PWNe \citep{Zhang08, Gelfand09,Tanaka10,Fang10,Bucciantini10}. 
Some of these models are applied to 
explain multi-wavelength spectra of a number of GeV or TeV PWNe (e.g. Crab Nebula, MSH 15$-$52, HESS J1825$-$137, 
Vela X, PWN G0.9+0.1 and G338.3+0.0) detected by ACTs and/or \fermi\ LAT, 
and obtain reasonably satisfactory results.
The accumulated particle spectra derived from essentially all of these PWNe can be described by broken power-laws
with the break energies lying around $100-500$~GeV.
\citet{Tanaka10} used the broadband spectrum of the Crab Nebula to calibrate their 
evolution model, and found that a fast decaying magnetic field ($\sim t^{-1.5}$) and 
a slowly evolving spectrum of the accumulated leptons (except for the first 1000~yr or so when $B$ is still large)
can possibly explain the trend of increasing 
TeV gamma-rays to X-ray ratio with age, as implied by recent multi-wavelength 
observations of PWNe. This weak time-dependence of the accumulated particle spectrum in PWNe has also been predicted
by some other similar evolution model \citep[e.g.,][]{Fang10}. 
In Table~\ref{pwntable}, we list out the spectral parameters for the accumulated particles predicted by a number of successful 
spectral evolution models for 5 representative gamma-ray PWNe observed in the TeV band. 
Among them, 4 of which also have firm detections by \fermi\ LAT.
For all models, we choose the parameters for $t > 1000$~yr
when the particle spectra become stable and are subjected to only weak time-dependence for the rest of the PWN lifetime.  

\begin{deluxetable}{lcccccccc}
\tablecolumns{9}
\tablewidth{0pt}
\tabletypesize{\footnotesize}
\tablecaption{Accumulated lepton spectra in 5 PWNe detected by \fermi\ and ACTs}
\tablehead{
\colhead{Source} & \colhead{Age}	& \colhead{Distance} & \multicolumn{2}{c}{Firm detection}
& \multicolumn{3}{c}{Spectral parameters} & \colhead{Refs.} \\\\\cline{4-5} \cline{6-8}
\multicolumn{9}{c}{} \\
\colhead{Name}	& \colhead{kyr}	& \colhead{kpc} & \colhead{ACTs} & \colhead{\fermi} & 
\colhead{$E_{br}$~[GeV]} & \colhead{$\alpha_{LE}$} & \colhead{$\alpha_{HE}$} & \colhead{}\\
}
\startdata
Crab & $1.0$ & $2.0$ & Y & Y & $300$ & $1.5$ & $3.1$  & \tnb \\ 
MSH~15-52 & $1.7$ & $5.0$ & Y & Y & $460$ & $1.5$ & $2.9$ &  \tnc \\  
HESS~J1825-137 &$21$ & $3.9$ & Y & N & $120$ & $1.0$ & $3.0$ &  \tnd \\ 
G0.9+0.1 & $6.5$ & $8.5$ & Y & N & $80$ & $1.7$ & $3.3$ & \tne \\ 
G338.3-0.0 & $4.5$ & $10.0$ & Y & Y & $260$ & $1.0$ & $2.8$ & \tnf \\
\enddata
\label{pwntable}
\tablenotetext{a}{Effective spectra of the accelerated and accumulated electrons and positrons inside the PWNe, 
approximated by broken power-laws. 
For each source, if multiple references are given, the spectral model makes reference to the first listed.}
\tablenotetext{b}{\citet{Tanaka10,Abdo10crab}} 
\tablenotetext{c}{\citet{Abdo10msh,Aharonian05a}} 
\tablenotetext{d}{\citet{Zhang08,Aharonian05c,Aharonian06e,Grondin11hess1825}} 
\tablenotetext{e}{\citet{Qiao09}}
\tablenotetext{f}{\citet{Fang10,Slane10}}
\end{deluxetable}

The actual process through which the accumulated energetic
particles are released from the PWNe into the Galactic space is known very little, mainly because
it is difficult to follow the evolution of a PWN lifetime through observation, and also that the available samples of Galactic PWNe at
different ages are only a handful. 
In the real situation, the real-time spectral shape of the `escaped' particles may differ from that of the 
`snapshot', effective spectra of the accumulated particles inside the PWNe directly derived by multi-wavelength modeling.
However, the accelerated particles are expected to be completely released into the ISM within $\sim 50$~kyr, 
the typical lifetime of PWNe.
As already mentioned above, spectral evolution models for PWNe have shown that time-evolution of the accumulated lepton spectrum inside the nebulae is expected to be very slow only for most part of a PWN lifetime, due to a fast-decaying magnetic field.
In the context of a straightly steady-state picture of our Galaxy, which we consider in this study, 
this subtle time-dependence within such short time-scale is unimportant. 
We hence consider it a reasonable assumption that the snapshot momentum distribution of the accumulated
particles can approximate to the first order the steady-state spectral shape of 
the $e^{+/-}$ pairs being injected into the ISM by the Galactic PWN ensemble. We note that a similar prediction has been made before the \pamela\ positron fraction measurement by \citet{Zhang01}.

\begin{figure}
\centering
\includegraphics[width=10cm]{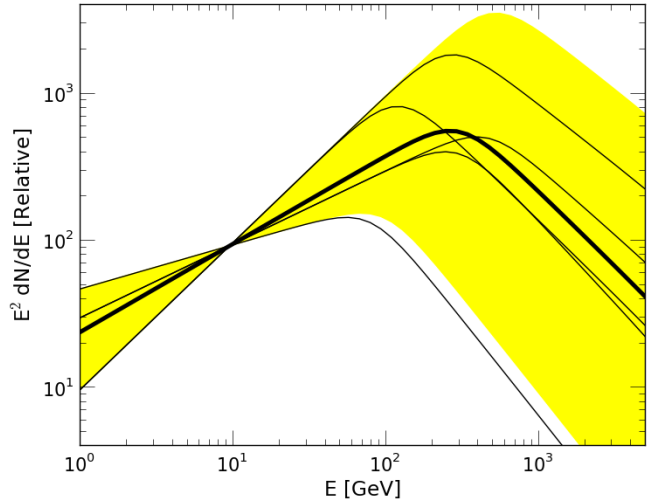}
\caption[Spectra of accumulated $e^+/e^-$ populations for 5 gamma-ray PWNe]
{Spectra of accumulated lepton populations deduced from broadband emission modelings for the 5 PWNe listed in Table~\ref{pwntable} 
(thin lines). All spectra are normalized to each other at 10~GeV for comparison.
The yellow band shows the sampled possible range of steady-state injection spectral shape for
leptons injected by the `PWN-propagation' source class.
The `mean' spectral shape from the sampled band is shown by the thick line.} 
\label{figpwninjectspectra}
\end{figure}

To determine this effective spectral shape, we refer to the measurement data and fitted parameters for the PWN sample given in Table~\ref{pwntable} for guidance. We adopt a broken power-law distribution, and allow the spectral parameters to vary over 
conservative ranges spanned by the underlying particle spectra found from these 5 gamma-ray PWNe. 
Specifically, we vary the break energy over the $100 - 500$~GeV band, 
and the pre-break and post-break power-law indices are sampled over values of $1.0-1.7$ and $2.8-3.3$, respectively. 
In Fig.~\ref{figpwninjectspectra}, we show the underlying spectra derived for the 5 gamma-ray PWNe with thin lines.
The range of steady-state injection spectrum we consider for the PWN-propagation class is shown with the yellow shaded band. 
The thick line is the `mean' spectral shape derived from this range of possible injection 
spectral shapes, which we will use as the reference spectrum (hereafter the `mean spectrum').
The steady-state CR injection luminosity for the Galactic PWN ensemble is determined jointly with the K$_{ep}$ parameter 
of the `SNR-propagation' class 
by normalizing to the absolute lepton flux and reproduce the $e^+ + e^-$ spectrum in the $50-150$~GeV band measured by \fermi\ 
\citep{Abdo09electron,Ackermann10electron}. This procedure is sufficient
to uniquely determine the absolute normalizations for both the `SNR-propagation' and `PWN-propagation' source classes.
In the results of this study (Fig.~\ref{figelectronpositron}, \ref{figposfraction} and \ref{fignewsource}), 
we will use a gray band and a thick solid line to show the models which adopt the range of possible spectrum and the mean spectrum 
for the PWN-propagation CR injection class, respectively. The gray band will be considered as the systematic uncertainty of our result
due to the uncertainty of the steady-state injection spectrum of the Galactic PWN ensemble.   

\section{Results and Discussion}
\label{results}
The primary and secondary CR spectra obtained for the three source classes 
described in section~\ref{simulations} are summed and compared
with the collection of available observations. 
In the figures, we show the calculated results using the default model parameters 
by the solid black lines, and their estimated systematic uncertainties 
by gray bands. The sensitivity of the results on the alternative setups $-$ 
different Galaxy halo sizes and an alternative source distribution $-$ 
will be discussed below.

\subsection{Anti-proton spectrum and fraction} 
\begin{figure}
\centering
\includegraphics[width=10cm]{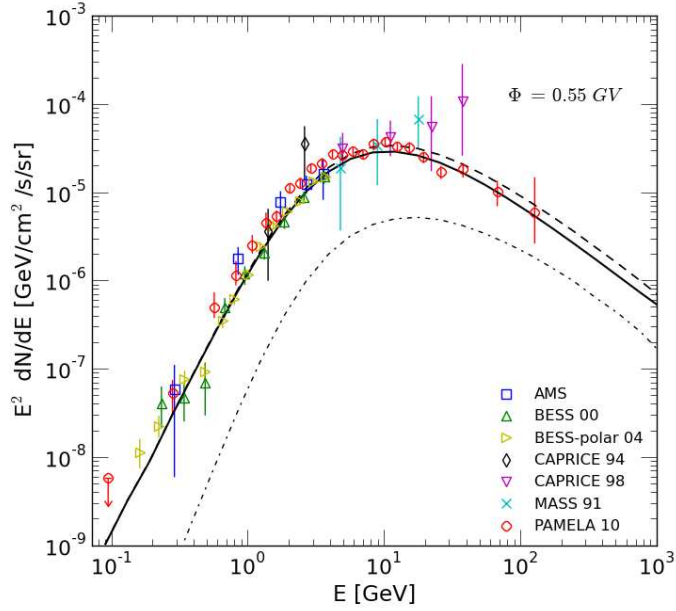}
\caption[Comparison of measured and calculated anti-proton spectra at Earth]
{Measured CR anti-proton spectra compared with
the default model. 
Anti-protons from the `SNR-propagation' class  
is shown by the solid line. The maximum possible contribution of $\bar{p}$ from the SNR-cloud interaction sites (upper-limit) 
is shown by the dash-dotted line. The dashed line is the sum of the two components.}
\label{figpbar}
\end{figure}

\begin{figure}
\centering
\includegraphics[width=10cm]{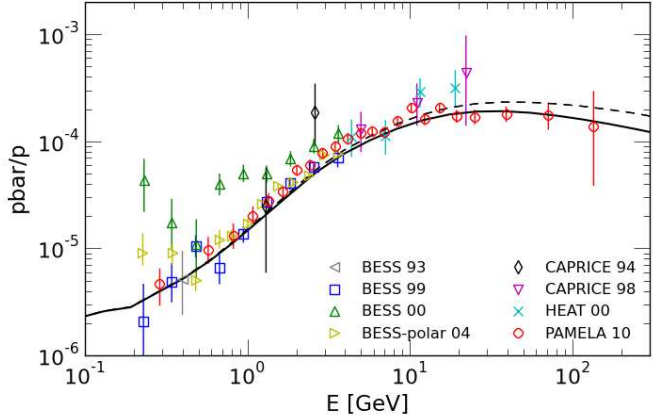} 
\caption[Comparison of measured and calculated Antiproton-to-proton ratio at Earth]
{Antiproton-to-proton ratio calculated for the default model 
compared with measurements. The solid line 
shows the ratio from the `SNR-propagation' class only. 
The dash line shows the ratio including also  
the maximum contribution from the SNR-cloud interaction sites (dash-dotted line in Fig.~\ref{figpbar}).} 
\label{figantiproton}
\end{figure}

The $\bar{p}$ spectrum and $\bar{p}/p$ 
ratio predicted by the model are found to be consistent with measurements by \pamela\ 
and other experiments in Figs.~\ref{figpbar} and \ref{figantiproton} respectively. 
The `SNR-cloud interaction' class contributes in addition
to the `SNR-propagation' component to the $\bar{p}$ intensity with a
similar spectral shape. 
Using the latest measurements of $\bar{p}$ spectrum by \pamela, especially above $10$~GeV,
it is possible to obtain a crude upper-limit on the total number of SNRs interacting with molecular clouds in
the Galaxy by requiring that the total model $\bar{p}$ spectrum does not overshoot the \pamela\ data points and their errorbars. 
Given the assumed source distribution, underlying proton spectrum, and typical $n \times W_p$ (see Section~\ref{Secondary}), 
this upper-limit is estimated to be $\sim 200$.
In Figs.~\ref{figpbar} and \ref{figantiproton}, the maximum possible contributions from the interaction sites 
are shown for illustration. Of course, the typical value of $n \times W_p$ adopted is derived from a rather small
sample observed by \fermi, meaning a possible significant systematic error on the upper-limit for the total number of these
objects in our Galaxy. However, their maximum contribution of secondary CR estimated basing on $\bar{p}$ data is independent of 
the exact value of $n \times W_p$ used, and should be reasonably robust.   
Future high-precision measurements of the $\bar{p}$ spectrum and $\bar{p}/p$ 
ratio, as well as gamma-ray observations of more SNR-cloud interaction systems, 
are expected to provide further constraints on the contributions from these secondary CR injectors.      

\subsection{Electron $+$ positron (lepton) spectrum}
\begin{figure}
\centering
\includegraphics[width=12cm]{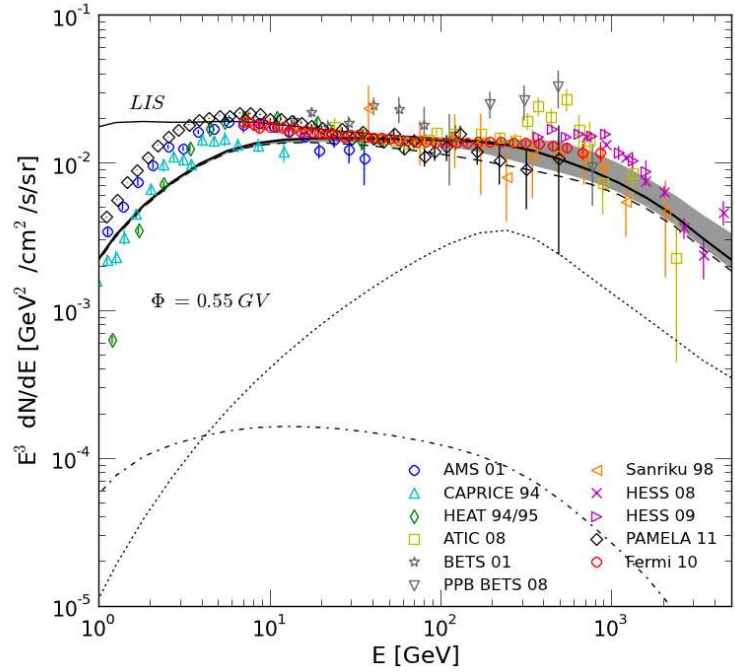}
\caption[Comparison of measured and calculated $e^+ + e^-$ spectrum at Earth]
{Measured CR lepton spectrum compared with the default model 
(thick solid). The thin solid line shows the 
LIS without solar modulation. 
Contributions from the `SNR-propagation' and  
`PWN-propagation' source classes  
are shown by the dashed and dotted lines respectively.  Note that the error bars 
for the \pamela\ points are statistical only. 
The dash-dotted line show the maximum contribution from the SNR-cloud interaction sites. 
The gray band shows the estimated systematic uncertainty of 
the total spectrum (see section~\ref{PWNinjection} for detail).} 
\label{figelectronpositron}
\end{figure}

We see in Fig.~\ref{figelectronpositron} that 
our model agrees with the observed lepton spectrum reasonably well above $\sim 20$~GeV, 
within the estimated systematic uncertainty. 
A majority of the CR lepton flux is made up by the SNR-injected 
primary $e^-$ which, however, comes short of reproducing the observation data 
at higher energies; in particular, the points measured by \fermi\  \citep{Ackermann10electron} 
and \pamela\ \citep{Adriani11a}. 
The PWN-injected leptons fill up this gap and dominate in the higher energies,
bringing the total spectrum back to agreement with the observed flux.
The key for the reproduction of the lepton data is 
the break energy of the PWN injection 
spectrum at a few hundred GeV. 
Contribution from the SNR-cloud interaction sites is constrained by the upper-limit estimated above from
the $\bar{p}$ spectrum, and is found to be almost negligible relative to the total spectrum at all energies.
The maximum secondary lepton flux from the SNR-cloud interaction sites is also shown in Fig.~\ref{figelectronpositron}. 
In terms of the total lepton energy budget in the Galaxy above $1$~GeV, 
contributions from `PWN-propagation' class and SNR-cloud interaction sites
make up for only about $2$~\%, while the `SNR-propagation' class
is responsible for the remaining $98$~\%. 

Below $\sim 20$~GeV, discrepancy is found between 
the measured lepton spectrum and the model. 
CR spectra in this energy regime are known to be strongly affected by the heliospheric 
modulation \citep[e.g., ][]{Shikaze07}, and the inaccuracy of the force-field 
approximation applied
\citep{Gleeson68} can be partially responsible for the disagreement. 
We do not observe, however, such a large discrepancy with observation 
in the CR $p$ and $\bar{p}$ spectra, 
which would have been affected by a similar modulation effect. 
Although we can not rule out completely the modulation effect as the cause 
of this discrepancy, we will explore an alternative possibility later.   

\subsection{Positron fraction}
\begin{figure}
\centering
\includegraphics[width=12cm]{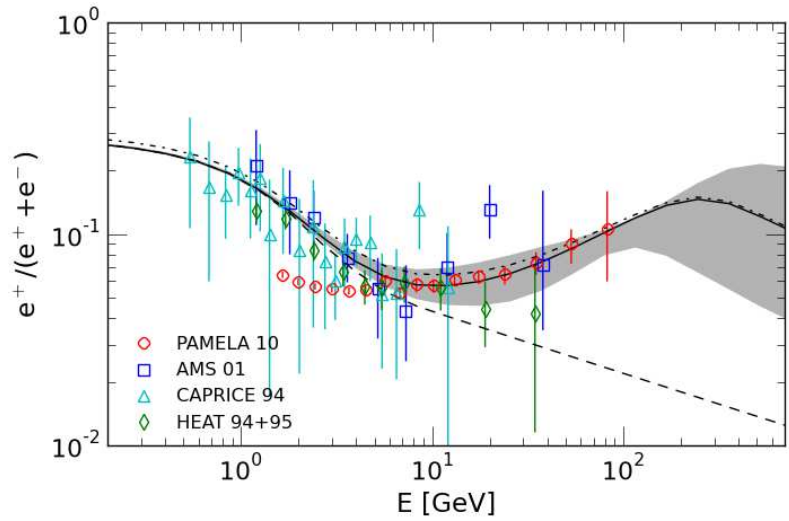}
\caption[Comparison of measured and calculated $e^+/(e^- + e^+)$ ratio at Earth]
{Measured positron fraction compared with that calculated using the default model. 
The fraction calculated for the `SNR-propagation' class alone is shown 
by the dashed line. The solid line shows the total of the `SNR-propagation' and 'PWN-propagation' components, 
with the gray band representing the associated systematic uncertainty. 
The dash-dotted line adds the maximum contribution from the SNR-cloud interaction sites 
(i.e. the dash-dotted line in Fig.~\ref{figelectronpositron}) to the solid line, 
showing the possible effect of secondary CR injection from these objects on the observed positron fraction.}
\label{figposfraction}
\end{figure}

One general feature predicted by our model is an enhancement 
of the positron fraction above $\sim 10$~GeV (Fig.~\ref{figposfraction}), 
as was recently observed by the \pamela\ experiment. 
This enhancement is found to be closely related to the hard broken 
power-law spectrum of the PWN-injected leptons. Without the $e^+$ contribution 
from the Galactic PWN ensemble, the `SNR-propagation' source class 
predicts a strictly decreasing  positron fraction with energy 
in contradiction to the \pamela\ points. Injection of secondary
$e^+$ from the SNR-cloud interaction sites may boost the positron 
fraction by $< 15$~\% for
all energies, but is obviously insufficient to reproduce the `enhancement' implied 
by the \pamela\ data. 

The fractions measured by \pamela\
in the $1-4$~GeV energy range is systematically lower than previous measurements. 
One possible reason is a charge sign-dependent heliospheric modulation effect. It has long
been proposed that the solar modulation of the Galactic CR propagating to the Earth depends on the particle charge and mass. The dependency changes significantly as the polarity of the solar magnetic field switches \citep[e.g.,][]{Clem96,Boella01,Clem09}. 
A significant drop of the positron ratio is indeed predicted during the first few years of the \pamela\ mission time when the Sun is in its A$^-$ minimum phase \citep{Picozza11}.  
The $AMS-02$ team has studied the energy dependence of the solar modulation to predict the positron ratio to be reduced significantly below $\sim 5$~GeV \citep{Bobik10a,Bobik10b}.  We reserve a more careful treatment            
of the solar modulation effect for future study.  

On the other hand, the large systematic uncertainty estimated above $100$~GeV mainly arises
from the poorly constrained break energy of the steady-state PWN injection spectrum, which is seen to vary roughly 
in the broad range of $100-500$~GeV from gamma-ray observations. 

We found that positrons injected by the Galactic ensemble of PWNe play a central role in 
reproducing the observed positron fraction enhancement. There is however a possibility that 
only PWNe lying within a few 100~pc from Earth are significant in contributing to the $e^+$ flux.
We use the propagation code described in the Appendix to estimate this local contribution from sources
with distance-to-Earth smaller than 300~pc, with the same continuous source distribution 
adopted in our default GALPROP model. The calculation shows that these nearby PWNe 
make up less than $10\%$ of the total calculated $e^+$ spectrum from the `PWN-propagation' source class in the \pamela\ energy band.  In 
the steady-state picture of Galactic CR injection studied in this work, PWNe beyond the local space  
explain most of the observed $e^+$ spectrum and fraction. This does not, however, 
rule out the possibility that a small fraction (up to $\sim 30$~\%) of the enhancement observed by \pamela\ 
is due to \textit{non-steady-state} injection from a few nearby PWNe, as already suggested in 
some recent papers \citep[e.g.,][]{Grasso09}.    

\subsection{A possible low-energy electron component}

\begin{figure}
\centering
\begin{tabular}{c}
\epsfig{file=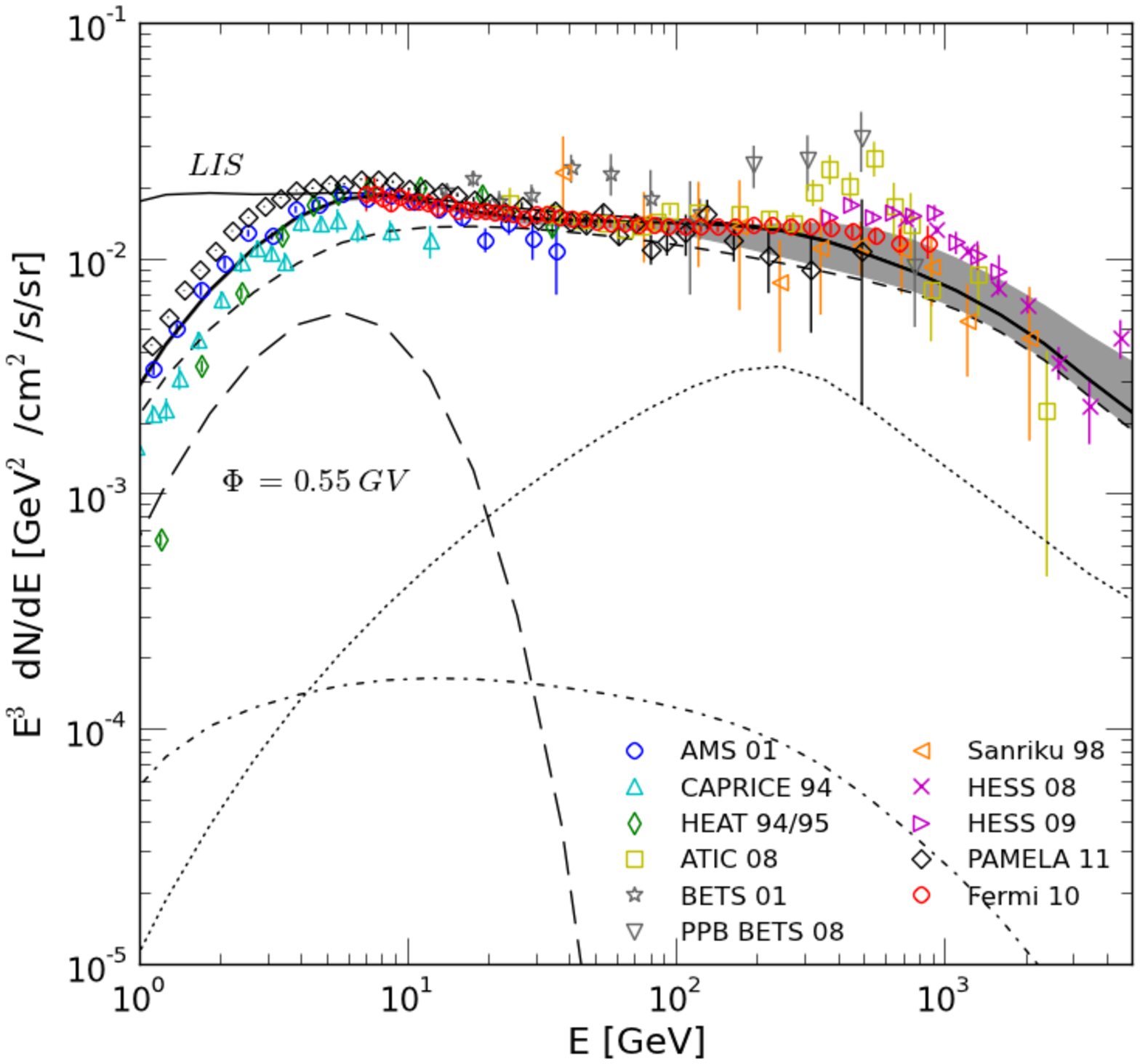,width=10cm}\\
\epsfig{file=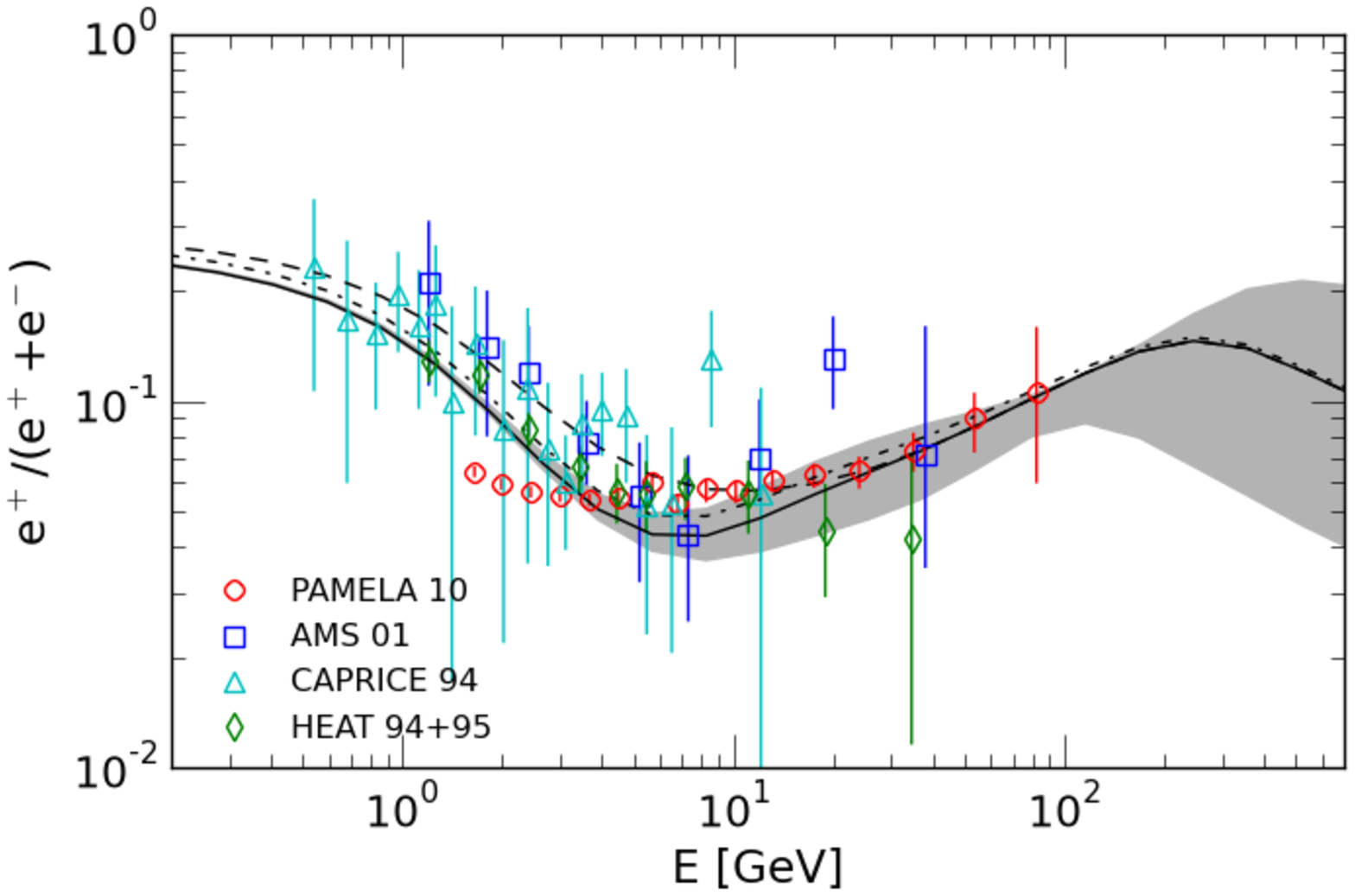,width=10cm}
\end{tabular}
\caption[Effect of speculative low-energy $e^-$ component on the calculated $e^+/(e^- + e^+)$ ratio at Earth]
{Top panel: Measured CR lepton spectrum compared with the 
default model (thick solid) including a possible new electron component in the low energy band
(long dash). Other lines are the identical to those in Fig.~\ref{figelectronpositron}. 
Bottom panel: Measured positron fraction compared with that calculated using the default model. 
The dashed line is the same as the solid line in Fig.~\ref{figposfraction}, while the solid line now
shows the modified positron fraction taking into account the assumed new electron component. 
The systematic uncertainty associated with the modified ratio is displayed by the gray band. 
The dash-dotted line shows the positron fraction when the maximum contribution from the SNR-cloud interaction sites 
is also included.}
\label{fignewsource}
\end{figure}

A possible cause for the low-energy discrepancy of the lepton 
spectrum is that a class of CR sources (other than the presumed Galactic ensemble of 
SNRs and PWNe) may exist within a few 100~pc of Earth that contribute 
predominately low-energy $e^-$, but not $e^+$. We note that 
\Fermi\ has found gamma-ray emission from X-ray binaries \citep{Abdo09LS61,
Abdo09LS5039} and $Suzaku$ has found gamma-ray pulsar-like periodic 
hard X-ray emission from a white dwarf \citep{Terada08} suggesting acceleration 
of electrons in its magnetosphere. 
Assuming the existence of such low-energy $e^-$ sources, 
we add an ad-hoc power-law spectrum with an exponential cutoff 
to the calculated lepton spectrum 
as shown in the upper panel of Fig.~\ref{fignewsource}. 
The positron fraction is naturally reduced accordingly as shown in the bottom panel. 
Since CR $e^-$ with energy $\sim 10$~GeV propagate only 
a few 100~pc over 1~Myr, these speculative sources should be relatively close to Earth.  
However, it is in order to note that this low-energy discrepancy is a result of one of our assumptions
that the injected protons and primary $e^-$ from SNRs share a common injection spectral shape
(for energies below $E_\mrm{max}$ of $e^-$), which is supported by the standard DSA theory as well as on general physics ground but have no direct observational evidence so far. 

\subsection{Gamma-ray spectrum}

\begin{figure}
\centering
\includegraphics[width=12cm]{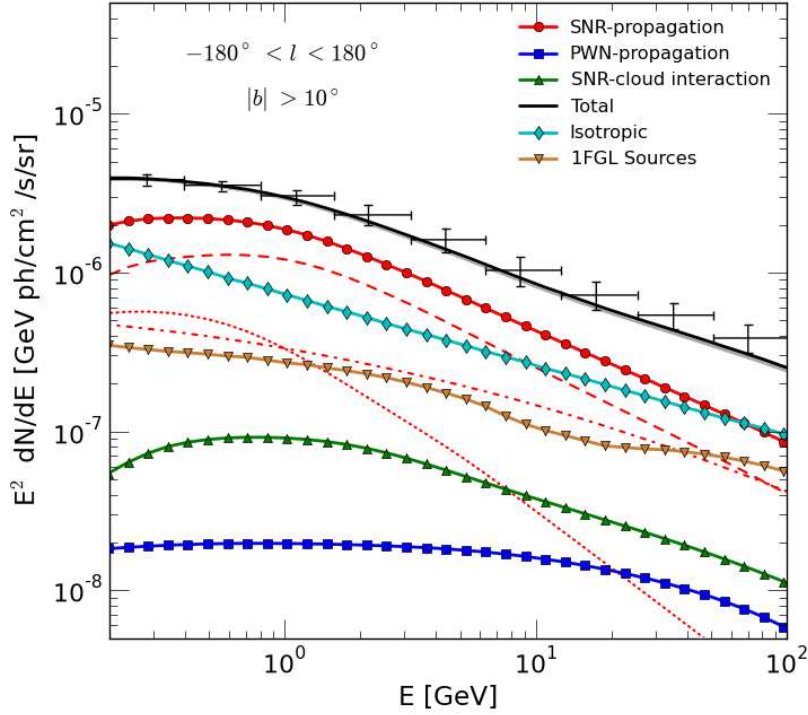}
\caption[Comparison of the calculated gamma-ray spectrum and that observed by \fermi\ LAT for $|b|>10^\circ$]  
{Gamma-ray spectrum observed by \fermi\ from the $|b|>10^\circ$  
region (points with error bars) compared with the default model. The total spectrum (solid black) 
includes the emission produced by SNR-injected CRs (red circle) and 
PWN-injected CRs (blue square) during their propagation, 
the emission from the SNR-cloud interaction sites (upperlimit) through pion-decay (green triangle), 
the isotropic background (cyan diamond), and 
the sum over the 1FGL \fermi\ catalog point-sources (brown inverted triangle). 
The emission represented by the red circle consists of 
pion-decay (dashed red), IC (dot-dashed red) and bremsstrahlung (dotted red) contributions. The emission produced by PWN-injected 
CRs consists predominantly of IC. The \fermi\ data points are taken from \citet{Abdo10isotropic}.}
\label{figgammaray}
\end{figure}

We compare in 
Fig.~\ref{figgammaray} the gamma-ray spectrum
calculated by our model for the $|b|>10^\circ$ region in the Galaxy 
with that measured by \fermi\ \citep{Abdo10isotropic}, 
which serves as an additional consistency check.
The contributions from the 3 source classes described 
in section~\ref{simulations} are shown separately. 
The total spectrum integrating over all source classes, gamma-ray point sources
from the 1st Fermi Catalog \citep{Abdo10catalog}, and the isotropic background 
(extragalactic diffuse emission and noise induced by CRs) comes out consistent with 
the data within error bars.\footnote{It is not surprising that the gamma-ray spectrum 
our model predicts is consistent with the \Fermi\ observation because 
the GALPROP proton injection spectrum has already been tuned to reproduce 
the Galactic gamma-ray spectrum.  More rigorous tests will be done when 
high quality analyses of large-scale Galactic diffuse emission become available.}   
The majority of the observed flux is dominated by the
diffuse gamma-rays produced by SNR-injected CRs along their propagation path,
the isotropic background, and the point sources. 
Diffuse emission produced by the PWN-injected 
leptons are relatively small. Within the upper-limit set for the total number of SNRs interacting clouds using
high-energy $\bar{p}$ data, gamma-rays emission from the `SNR-cloud interaction' class is found to be insignificant
in this region of the sky (the maximum contribution from the interaction sites are shown in the figure for illustration). 

\subsection{Effects of modified source distribution model and Galaxy halo heights}
Calculations performed for the default model are repeated for
the three alternative models - one with an alternative source distribution (see section~\ref{snrpwndistribution}), and the others 
with different Galaxy halo sizes. 
The impacts of the alternative setups on our results are summarized below: 
\begin{figure}
\centering
\includegraphics[width=12cm]{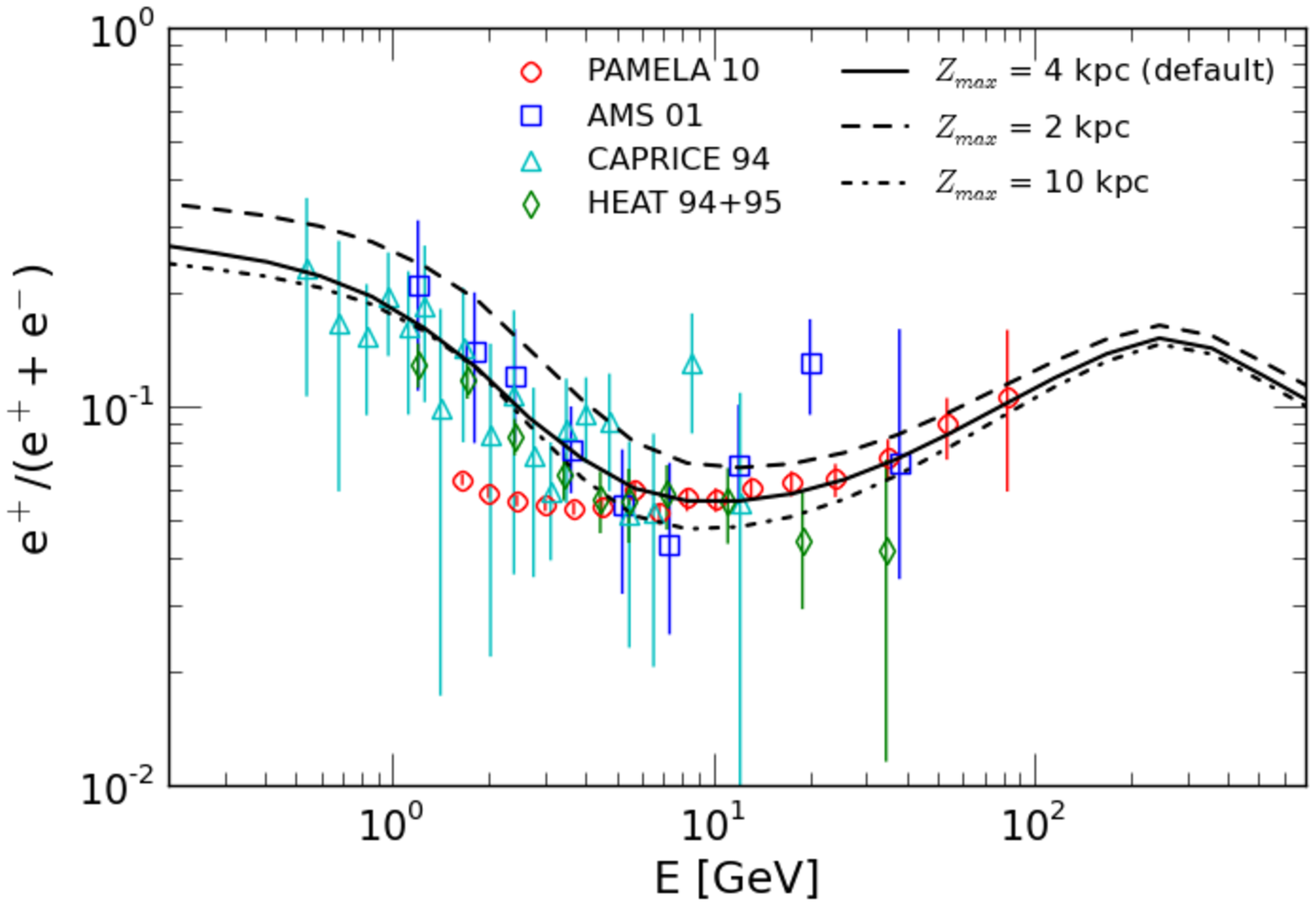}
\caption[Effect of Galaxy halo heights on the calculated positron ratio]  
{Comparison of positron fraction calculated using models with different assumed Galaxy halo heights. The mean spectrum
is used here for the PWN-propagation CR injection class for all models, without showing their associated systematic uncertainty
for clarity reason. The possible contribution from the SNR-cloud interaction sites is not included here.}
\label{figposfractionhalos}
\end{figure}

\begin{itemize}
\item To re-fit to the CR measurement data, the injection luminosities of primary CRs for both the `SNR-propagation' 
and `PWN-propagation' classes have to be slightly varied against those adopted 
for the default setup. The required adjustment in the injection luminosities, 
as well as the resulting total CR energy budget in the Galaxy, are 
summarized in Table~\ref{CRtable}. Most of the deviations are found to stay within $\pm 20$~\% from the default model. 

\item For models with different Galaxy halo sizes, the diffusion coefficient
has to be modified to reproduce the observed B/C ratio.
The change made is shown in Table~\ref{modelparamtable1}.

\item Our results are found to be essentially insensitive to the change of the assumed source distribution models, besides the
slight changes in CR injection luminosities needed. 

\item We notice that modification of the assumed Galaxy halo heights have subtle influence on the calculated positron fraction 
(Fig.~\ref{figposfractionhalos}). A smaller halo size is found to result in a slightly larger predicted ratio, and vice versa. 
The energy dependence of the ratio is also somewhat affected. With the consideration of the spread 
among the measurement data as well as the systematic uncertainty associated to the calculated fraction 
which stems from the assumed PWN injection spectrum, however, the observed change in the results 
are relatively insignificant for any strong conclusion to be drawn. The two models with altered
halo heights can still be considered to be in agreement with measurements.  

\item The gamma-ray spectrum shown in Fig.~\ref{figgammaray} for the $|b| > 10^\circ$ sky region 
does not alter in any significant way among the models. Models with different halo sizes changes the 
overall normalization of the calculated spectrum, but by only less than 10\% from the default model. 
More detailed comparison with large-scale gamma-ray data in the future, 
e.g. spectra from the Galactic ridge and the inner-Galaxy region,
as well as small-scale structures, may make it possible to discriminate against the models
and help put constraints on important quantities such as the Galaxy halo height, 
and the gas and source distributions in our Galaxy.
   
\end{itemize} 

\noindent At this point, we can argue that our results and conclusions obtained for the default model are robust against modifications 
of the source distribution and Galaxy halo heights.  

\begin{deluxetable}{lcccc}
\tablecolumns{5}
\tablewidth{0pt}
\tabletypesize{\footnotesize}
\tablecaption{Energetics of Galactic Cosmic Rays in the Steady State \tna}
\tablehead{
\colhead{Particle} & \colhead{$T_p>0.1$~($1.0$)~GeV}	& \colhead{Alt. Model 1\tnb}	& \colhead{Alt. Model 2\tnc}
&  \colhead{Alt. Model 3\tnd} \\
\colhead{} & \colhead{[erg] or [erg~$s^{-1}$]}	& \colhead{$\Delta$ in \%}	& \colhead{$\Delta$ in \%} & \colhead{$\Delta$ in \%}
}
\startdata
\sidehead{~~K.E. of CRs injected by SNRs integrated in the steady-state Galaxy \tne}
\hline
\\
$p$ & $7.5\times 10^{55}$ ($6.1\times 10^{55}$) & $+14.8$~\% & $...$ & $...$\\
$e^-$ (pri) & $1.9\times 10^{54}$ ($7.4\times 10^{53}$)	& $+9.3$~\% & $...$ & $...$	 \\
$e^-$ (sec) & $1.7\times 10^{53}$ ($5.2\times 10^{52}$) & $+1.8$~\% & $...$ & $...$	 \\
$e^+$ & $6.9\times 10^{53}$ ($1.5\times 10^{53}$) & $+1.5$~\% & $...$ & $...$	\\
$\bar{\mrm{p}}$ & $4.4\times 10^{51}$ ($4.4\times 10^{51}$)	& $-0.5$~\% & $...$ & $...$	 \\
\\
\hline
\sidehead{~~Injection luminosity of CRs from SNRs}
\hline
\\
$p$ & $6.4\times 10^{40}$ ($4.2\times 10^{40}$) & $+14.6$~\% & $+3.0$~\% & $-6.5$~\% \\
$e^-$ (pri) & $1.2\times 10^{39}$ ($8.0\times 10^{38}$) & $+4.1$~\% & $-23.3$~\% & $+21.1$~\% \\
\\
\hline
\sidehead{~~K.E. of CRs injected by PWNe integrated in the steady-state Galaxy \tnf}
\hline
\\
$e^-$ & $6.6\times 10^{51}$ ($4.8\times 10^{51}$) & $+9.2$~\% & $...$ & $...$ \\
$e^+$ & $6.6\times 10^{51}$ ($4.8\times 10^{51}$) & $+9.2$~\% & $...$ & $...$ \\
\\
\hline
\sidehead{~~Injection luminosity of CRs from PWNe \tnf}
\hline
\\
$e^-$ & $2.3\times 10^{37}$ ($2.2\times 10^{37}$) & $+0.7$~\% & $-19.0$~\% & $+19.1$~\% \\
$e^+$ & $2.3\times 10^{37}$ ($2.2\times 10^{37}$) & $+0.7$~\% & $-19.0$~\% & $+19.1$~\% \\
\enddata
\label{CRtable}
\tablenotetext{a}{The results listed here are extracted from the GALPROP outputs for our models.} 
\tablenotetext{b}{Model using the alternative source distribution (see Section~\ref{snrpwndistribution})}
\tablenotetext{c}{Model using a smaller Galactic halo height of $Z\mrm{_{max}}=2$~kpc.}
\tablenotetext{d}{Model using a larger Galactic halo height of $Z\mrm{_{max}}=10$~kpc.}
\tablenotetext{e}{Kinetic energy of the accumulated CRs within the assumed boundary of the Galaxy. 
Models with different propagation volumes are not compared.}
\tablenotetext{f}{Kinetic energy and luminosity correspond to the model using the mean spectrum for the steady-state injection
(see Section~\ref{PWNinjection}).}  
\end{deluxetable} 

\begin{deluxetable}{lccccc}
\tablecolumns{6}
\tablewidth{0pt}
\tabletypesize{\scriptsize}
\tablecaption{$\pi^0$-decay gamma-ray emissivities in different regions of the Galaxy \tna}
\tablehead{
\colhead{Region} & \colhead{$E_\gamma >0.1$~GeV} & \colhead{$E_{\gamma}>1.0$~GeV} & \colhead{Alt. Model I\tnc} & \colhead{Alt. Model II\tnd} &  \colhead{Alt. Model III\tne} \\
\colhead{} & \colhead{[s$^{-1}$H$^{-1}$]} & \colhead{[s$^{-1}$H$^{-1}$]} & \colhead{$\Delta$ in \%}	& \colhead{$\Delta$ in \%} & \colhead{$\Delta$ in \%}
}
\startdata
Inner-Galaxy\tna & $2.7\times 10^{-25}$ & $3.2\times 10^{-26}$ & $-29.7$~\% & $-29.3$~\% & $-13.5$~\% \\
Hi-latitude\tnb & $1.6\times 10^{-26}$ & $1.9\times 10^{-27}$ & $+10.2$~\% & $-5.1$~\% & $+43.3$~\% \\
Entire Galaxy    & $1.8\times 10^{-26}$ & $2.1\times 10^{-27}$ & $+9.1$~\% & $+5.1$~\% & $+33.1$~\% \\
\enddata
\label{Gammatable}
\tablenotetext{a}{$R < 5$~kpc, $|Z| < 500$~pc}
\tablenotetext{b}{$|Z| > 500$~pc}
\tablenotetext{c}{Model using the alternative source distribution (see Section~\ref{snrpwndistribution})}
\tablenotetext{d}{Model with Galaxy halo size of $Z\mrm{_{max}}=2$~kpc.}
\tablenotetext{e}{Model with Galaxy halo size of $Z\mrm{_{max}}=10$~kpc.}
\tablenotetext{f}{The results listed here are extracted from the GALPROP outputs for our models.} 
\end{deluxetable}  

\subsection{Energy budget of Galactic CR}
With gamma-ray observations of CR sources 
and measurements of charged CR spectra analyzed  
coherently under a common calculation platform using GALPROP, 
it is possible to extract a self-consistent set of quantities related to 
Galactic CRs, including the injected CR energies from SNRs and PWNe, 
interaction rate of nuclear CRs, and gamma-ray emissivity 
in the Galaxy. These quantities depend mildly on the 
presumed boundary of the propagation volume and the assumed 
source distribution. The results are summarized in Table~\ref{CRtable} and Table~\ref{Gammatable}. 
Based on the calculated energetics, various useful quantitative information can 
be extracted as we demonstrate below for the default set of model parameters: 

\begin{itemize}
\item From the total energy and luminosity given for primary protons 
in Table~\ref{CRtable}, we can roughly estimate the typical lifetime of CR protons ($T_p>0.1$~GeV) 
inside the Galaxy as $\sim 7.5\times 10^{55}/2.0\times 10^{48} = 3.8\times 10^7$~yr.
Similar estimation for CR leptons injected by PWNe is roughly 
$9.1\times 10^6$~yr, substantially shorter than the protons. 
This mean lifetime is roughly consistent with the synchrotron loss time-scale of electrons 
with an energy of $\sim 20$~GeV \citep[e.g.,][]{Yamazaki06}, close to the mean energy 
of leptons injected by the PWNe ensemble in our model. The difference in the lifetime becomes 
larger for CRs with $T > 1$~GeV: $\sim 4.6 \times 10^7$~yr for protons versus 
$\sim 6.9 \times 10^6$~yr for leptons. 

\item Approximating the CR proton spectrum in the Galaxy ($T_p>1$~GeV) by a broken PL shape 
consistent with LIS (E$_b=7$~GeV, $\gamma_1=2.0$, $\gamma_2=2.75$), we obtain the 
average kinetic energy per CR ($T_p>1$~GeV) to be 3.5~GeV or $5.6\times 10^{-3}$~erg. 

\item For a typical lifetime of $\sim 4 \times 10^7$~yr, protons cross a column 
density of $4n\times 10^{25}$cm$^{-2}$ where $n$ is the gas density 
averaged over the propagation volume bounded by $R_{max}=30$~kpc and 
$Z_{max}=4$~kpc. Since $pp$ inelastic cross-section is 
$\sim 30$~mb or $\sim 30\times 10^{-27}$~cm$^2$, 
about $1 \times n$ of CRs interact with the ISM gas. 
Since the mean kinetic energy of protons  is $\sim 3.5$~GeV, 
and $\pi^0$ multiplicity is $\sim 0.5$ \citep{Kamae05}. Hence a CR proton produces, 
on average, $\sim 1 \times n$ pionic gamma-rays in its life.


\item For a total number of target H-atom of $1.1 \times 10^{67}$ from H~I, 
H2 and H~II regions given in \citet{Ferriere01} and the pionic gamma-ray 
emissivity averaged over the default GALPROP cylindrical box,   
the number of pionic gamma-rays ($E_\gamma > 1$~GeV) coming 
out of the Galaxy can be estimated as 
$(4.5\times 10^{-27}) \times (1.1\times 10^{67}) = 5.0 \times 10^{40}$~s$^{-1}$. 
Since the total number of CR protons ($T_p>1$~GeV) is estimated to be
$6.1\times 10^{55}$/$5.6\times 10^{-3}$= $1.1\times 10^{58}$, the average number 
of pionic gamma-rays ($E_\gamma > 1$~GeV) is 
$(5.0 \times 10^{40}) / (1.1\times 10^{58}) = 5\times 10^{-18}$~s$^{-1}$ 
per CR proton, or $(10^{15}) \times (5\times 10^{-18}) = 5\times 10^{-3}$ 
integrated over the typical lifetime of one CR proton of $\sim 10^{15}$~s.
Hence, CR protons spend most of their life on average propagating in a low-density medium, 
with $n\sim 0.005 $~cm$^{-3}$, whereas the typical density 
of the `SNR-cloud interaction' sites is $\sim 100$~cm$^{-3}$.

\end{itemize}

Because of the constraints given by the gamma-ray observations of the 
CR sources and by the CR proton and lepton flux measurements, 
these estimations are stable within a $\sim 20$~\% level and provide a coherent picture 
of Galactic CR propagation. 

The overall success of the model in reproducing the observed CR ratios 
and spectra suggests that the majority of the CRs observed at Earth 
are steadily injected from the Galactic ensembles of SNRs and PWNe. 
The consistency between the observed and predicted diffuse gamma-ray spectra 
strengthens this interpretation. 

Dependence of the Galactic CR energy budget on CR diffusion models
has been studied by \citet{Strong10CR} within the GALPROP framework. 
They report how the diffusion coefficient varies for the 3 halo sizes 
assumed in the present study (2kpc, 4kpc and 10kpc), for the plain diffusion 
and for the diffusive re-acceleration models, under generic constrains that radio, \fermi\
electron \citep{Abdo09electron} and \fermi\ diffuse gamma-ray data
\citep{Abdo09localh1, Abdo09noexcess, Abdo10isotropic, Abdo10outergalaxy} 
to be reproduced. The CR injection spectra and source distribution have been 
fixed. Our diffusion coefficients and re-acceleration parameters corresponds 
to their Model2 (z04LMS).

By comparing the study by \citet{Strong10CR} with that presented here, we note that 
the CR injection luminosity is much less dependent on the halo size. This means 
Galactic CR modeling is constrained better by CR data and gamma-ray measurements 
near the region where injection is taking place. As the \fermi\ data accumulate further, 
we will be able to combine the two approach and constrain the diffusion process 
with measurements.

\section{Summary and Conclusions}
\label{conclusions}

The analysis presented here has brought the recent gamma-ray observations
of CR acceleration sites by \fermi\ and ACTs and the recent charged CR 
measurements at Earth to one common platform of GALPROP. 
We were able to do this by assuming all CR species in the Galaxy are in a steady state, 
on which GALPROP is built. CRs are assumed to be injected only 
from SNRs and PWNe with the spectra averaged over their respective 
evolution history. The sources are distributed, approximately proportional to, 
either distribution of pulsars or SNRs in the 
Galacto-centric cylindrical coordinate system. 
Throughout the analysis we have not used the positron fraction neither explicitly 
nor implicitly to constrain the input parameters for GALPROP. The analysis 
has also remained totally blind to the measured $\bar{p}/p$ ratio. 
With these assumptions, the model reproduces the positron fraction
observed by \pamela\ reasonably well (Fig.~\ref{figposfraction}), and 
the $\bar{p}/p$ ratio by \pamela\ (Fig.~\ref{figantiproton}) very well 
under constraint of the observed total $e^- + e^+$ spectrum (Fig.~\ref{figelectronpositron}) 
\footnote{We note that our model inherits the prediction of conventional GALPROP models on the 
$\bar{p}/p$ ratio except for the contribution from the SNR-cloud interaction.}.
We have tested the robustness of the model results against alternative assumptions on source distribution
as well as the Galaxy halo height, and confirmed that the conclusions drawn in this study stay intact.

We conclude that the energy dependence of the excess in positron fraction is mostly 
explained by the contribution from the Galactic ensemble of PWNe added upon the 
secondary $e^+$ produced by SNR-injected CR protons along their propagation path in the ISM. This does not, however, exclude contribution coming from local PWNe 
up to $\sim 30-40$~\% in flux.
The observed $\bar{p}$ flux is predominantly attributed to interactions of the primary 
nuclear CRs along their propagation paths, while those potentially injected from 
dense clouds interacting with middle-aged SNRs are estimated to be 
smaller than $\sim 11$~\% at most in total flux above $100$~MeV. 
The steady-state contribution from local SNRs and PWNe (within 300~pc from Earth) is 
estimated to be less than $\sim 10$~\% in the $e^+$ spectrum,
which is not crucial in any way for reproducing the observed positron fraction. 

Our result implies that for a typical $n \times W_p$ 
($T_p > 1$~GeV) of $5 \times 10^{51}$~erg/cm$^3$,
up to $\sim 200$ SNRs can be interacting with dense clouds and injecting secondary CRs (section~\ref{results}).
The maximum contribution of positrons at Earth from these interaction sites 
is found to be relatively insignificant and in itself cannot reproduce the observed positron fraction without
including injection from the PWNe.  

By assuming a SN rate in the Milky Way of one in 30 years~\footnote{
\citet{Diehl06} estimate the rate of core-collapse supernovae to be $\sim 1.9\pm 1.1$
per century. The Type Ia supernovae rate is less constrained observationally, ranging
from $0.3-1.1$ per century \citep[e.g.][]{Mannucci05,Knodlseder05}. These two
rates add up to $\sim 3$ per century.}, the injection luminosity 
of primary CR protons from SNRs listed in Table~\ref{CRtable}, 
and SNRs' active life of $5 \times 10^4$~yr, we can estimate that $\sim 1700$ 
SNRs are actively injecting CRs into the ISM, 
and each SNR injects $\sim 6.1 \times 10^{49}$~erg of CRs ($E>0.1$~GeV) into the 
Galactic space in its lifetime.
Similarly, if we assume a PWN birth rate of one in 50
years and that PWNe remain active for $10^5$~yr on average, $\sim 2000$ PWNe 
are injecting CR leptons, with each PWN injecting $\sim 7.2 \times 10^{46}$~erg of 
CR leptons into the Galactic space. CR protons with $T_p>1$~GeV stay 
$\sim 4.6 \times 10^7$~yr in the Galaxy 
with only $0.1$\% interacting with gas while CR leptons with $T_e>1$~GeV lose 
energy in $\sim 6.9 \times 10^6$~yr.   

A noticeable discrepancy with data is found in the low energy total lepton spectrum 
below $\sim 20$~GeV. If the corrections for the solar modulation effect 
are properly approximated to the measured spectra, 
it may imply that an unknown additional class of \textit{local} CR sources which 
preferentially inject low-energy ($E<20$~GeV) electrons is required. 

Higher precision gamma-ray spectra expected from \Fermi\ and \textit{CTA}, as well as 
precision measurements on the positron fraction above 100~GeV and
the lepton spectrum in the TeV range by \amsII\ \citep{AMSII09} will allow us to extract important 
information on the evolution of PWNe and SNRs. 
\Fermi\ will soon measure anisotropy in the gamma-ray emissivity at
nearby molecular clouds, and either detect or exclude the
proposed GRB scenario for the \pamela\ $e^+$ excess \citep{Fujita09, Perrot03}.
Future high precision measurement of $\bar{p}$ by \amsII\
will further constrain contribution from the SNR-cloud interaction sites.

\vspace{12pt}
\noindent\textbf{Acknowledgements}
\vspace{12pt}\\
We thank I. Moskalenko, L. Tibaldo, S. Digel, N. Omodei, and M. Ackermann for careful reading of the manuscript and suggestions for improvement. We have also benefited from discussion with members of the Fermi-LAT collaboration and suggestions by the anonymous referee of this journal. 
Our participation to the Fermi-LAT collaboration has been supported by the
Department of Energy in the United States, the Commissariat \`a l'Energie Atomique
and the Centre National de la Recherche Scientifique / Institut National de Physique
Nucl\'eaire et de Physique des Particules in France, the European Community, 
the Agenzia Spaziale Italiana
and the Istituto Nazionale di Fisica Nucleare in Italy, the Ministry of Education,
Culture, Sports, Science and Technology (MEXT), High Energy Accelerator Research
Organization (KEK) and Japan Aerospace Exploration Agency (JAXA) in Japan.

\appendix
\section{A Supplementary Propagation Scheme}
\label{appendix}
Models described in this study are primarily built using GALPROP as far as possible. 
Other calculations that cannot be readily implemented in a typical GALPROP model are accomplished via a  
simple propagation code separately developed by us, which include the following: 
(1) the local contribution of CR positrons from PWNe 
lying within a given distance from Earth; 
(2) contribution of secondary particles 
injected from SNRs interacting with clouds; and 
(3) the high-energy rollover inherit to the primary $e^-$ spectrum injected from SNRs due to 
spectral cutoff at a time-evolving maximum energy. To obtain the CR spectra 
propagated from a continuous source distribution characterized by distance-to-Earth $\mathbi{r}$, source age $t_\mrm{age}$, 
and the active lifetime of each source $t_\mrm{life}$ (time
interval during which CRs are injected into the ISM from a source),
we use the analytic solution of the simple diffusion-loss transport equation adapted from \citet{Atoyan95}
\footnote{This propagation calculation does not include re-acceleration effect 
and interaction involving heavy nuclei which can possibly affect the B/C ratio.}:

\begin{equation}
N(E,\mathbi{r}) = \int_{t_\mrm{age}-t_\mrm{life}}^{t_\mrm{age}}\frac{Q_\mrm{inj}(E_i,\mathbi{r},t')b(E_i)}{\pi^{3/2}b(E)r_\mrm{D}^3(E,t')}e^{-(|\mathbi{r}|/r_\mrm{D}(E,t'))^2}dt'
\label{eqn_transsolution}
\end{equation}

\begin{equation}
r_\mrm{D}(E,t) \approx 2\sqrt{D(E)\frac{1-(1-b_0Et)^{1-\delta}}{(1-\delta )b_0Et}}
\label{eqn_difflen}
\end{equation} 

\noindent Here $b(E)$ is the energy-loss rate of the CRs, which can be expressed as 
$b(E) \approx b_0E^2$ for synchrotron and IC losses of leptons during propagation, while 
loss is considered negligible for nuclei 
(such that equation~\ref{eqn_transsolution} 
and \ref{eqn_difflen} reduce to the usual solution of a simple 1-D diffusion equation
and diffusion length respectively). 
$E_i$ is the initial injection energy of CRs whose observed energy at Earth is $E$. 
The effective diffusion length $r_\mrm{D}(E,t)$ for a particle with observed energy $E$  
and propagation time $t$ is given by equation~\ref{eqn_difflen}, in which energy loss 
effect is accounted for. $D(E) \propto \beta E^{\delta}$ is the adopted spatial diffusion 
coefficient, where the index $\delta = 1/3$ is used in consistence with the GALPROP 
calculations. The CR injection term is given by $Q_\mrm{inj}(E,\mathbi{r},t) = 
Q(E,t)n_\mrm{soc}(\mathbi{r},t)$ where $Q(E,t)$ is the injection spectrum and 
$n_\mrm{soc}(\mathbi{r},t)$ is the source distribution in space-time. Spatial 
distribution of sources is the same as the distribution 
used in the default GALPROP model, while $t_\mrm{age}$ of the sources are assumed to be distributed 
uniformly in time starting from a maximum of $10^8$~yr up to $5 \times 10^4$~yr from now. 
For every source, $t_\mrm{life}$ is assumed to be $5 \times 10^4$~yr. 

For (1), the injection spectra and luminosities of $e^+$ and $e^-$ are directly 
taken from the GALPROP inputs and outputs adopted for the `PWN-propagation' class of the 
default model (see Tables~\ref{modelparamtable1} and \ref{CRtable}).
For (2), these are calculated for the secondary $e^+$, $e^-$ and $\bar{p}$ according to the description in Section~\ref{Secondary}. 
Tertiary CRs produced through hadronic interactions during propagation of $\bar{p}$ in the ISM are neglected. 
For (3), a time-dependent exponential cut-off is applied to the 
injection term for the SNR-injected primary $e^-$ ($Q(E,t) \propto e^{-E/E_\mrm{max}(t)}$) to incorporate the time-evolving
maximum electron energy $E_\mrm{max}(t)$ given by equation~\ref{eqn_emax}. This results 
into a smooth high-energy turnover of the propagated primary $e^-$ spectrum at Earth, which commits at around $1$~TeV.
This calculated turnover is then applied to the GALPROP-calculated primary $e^-$ spectrum from the `SNR-propagation'
source class. 

\bibliographystyle{model1-num-names}

\end{document}